\documentclass[%
 reprint,
 showpacs,preprintnumbers,
 amsmath,amssymb,
 aps,
 prl,
]{revtex4-1}
\usepackage{dcolumn}
\usepackage[dvipdfm]{graphicx}
\usepackage{subfigure,color}
\usepackage{mathrsfs}

\makeatletter
\def\btt#1{\texttt{\@backslashchar#1}}
\DeclareRobustCommand\bblash{\btt{\@backslashchar}} \makeatother

\begin{document}

\title{ Quantum Landau Spin Hall Insulator %in the presence of mirror symmetry
}

\author{Tetsuro Habe$^1$ and Yasuhiro Asano$^{1,2}$}
\affiliation{$^1$Department of Applied Physics,
Hokkaido University, Sapporo 060-8628, Japan}
\affiliation{$^2$Center for Topological Science \& Technology,
Hokkaido University, Sapporo 060-8628, Japan}

\date{\today}

\begin{abstract}
We study theoretically the two-dimensional topological electric state in a single band semiconductor with strong spin-orbit interactions
under harmonic scalar electrostatic potentials.
The electronic states described by the spin Landau levels are insulating in the bulk and host 
gapless edge states in the presence of time-reversal symmetry.
Such topological states show the properties of the quantized electric conductivity and the quantized spin Hall conductivity 
characterized by the spin Chern number $\mathbb{Z}$.
The quantization of the two conductivities are 
robust under various perturbations such as the potential disorder, the Zeeman field, and the spin-orbit scatterings.
Existing semiconductor technologies would realize the topological states discussed in the present paper.
\end{abstract}


\maketitle

The quantum spin Hall state (QSHS) is the two-dimensional time-reversal invariant topological electric state 
characterized by $\mathbb{Z}_2$ topological number~\cite{Kane2005b,Kane2005a,Moore2007,Roy2009a}.
In the topological phase, the non-trivial topological number $\mathbb{Z}_2=1$ suggests the presence of a pair of gapless 
conducting channels at the edge of the quantum spin Hall insulator~\cite{Qi2006}.
The time-reversal symmetry protects such edge channels because they are Kramers partner of each other.
The key features for realizing the $\mathbb{Z}_2$ QSHS are the band inversion between the conduction and valence bands, 
and the spin-orbit interaction reopening the gap in the energy spectra. 
Experimentally, the QSHS was first confirmed in a quantum well of HgTe sandwiched by CdTe according to the 
theoretical prediction~\cite{Bernevig2006,Bernevig2006b,Konig2007}.
However, the band inversion is possible only in wide enough quantum wells, where 
the spin Hall conductivity is not well quantized because of the imperfect two-dimensionality.
Recently, theoretical studies have suggested the $\mathbb{Z}_2$ QSHSs in silicene and germanene\cite{Liu2011,Ezawa2012,An2013}. 
These materials have the two-dimensional lattice structures similar to the graphene.
Generally speaking, the quantization of the spin Hall conductivity is not achieved in such QSHSs 
because the spin-orbit interaction tends to flip spin of an electron. 
Indeed in silicene and germanene, the deformation of the hexagonal lattice structure in the perpendicular to the plane 
breaks the mirror symmetry intrinsically~\cite{Vogt2012}, which causes the strong Rashba spin-orbit interaction 
and damages the quantization of the spin Hall conductivity.
In addition, it is still difficult to make the experimental setup for measuring transport properties 
in silicene and germanene at the present time.
Tunable topological states on fabricable materials by existing technology have been highly desired to make clear rich physics of topological materials in experiments.

In this paper, we propose an artificial quantum spin Hall insulator fabricated on a mirror symmetrical quantum 
well of a single-band semiconductor with strong spin-orbit interactions.
The proposed quantum spin Hall insulators have four novel features as follows. 
At first, the QSHSs are described by the Landau levels and are characterized by the spin Chern number $\mathbb{Z}$ which 
can be any integer numbers~\cite{Sheng2006}.
Therefore we call such topological state "quantum Landau spin Hall state" (QLSHS).
At the edge of the spin Hall insulator, $\mathbb{Z}$ pairs of Kramers gapless edge channels carry 
the spin polarized current. Secondly the spin Hall conductivity is quantized, which is a direct consequence of the spin Chern number.
Thirdly the spin polarized edge channels are robust under the symmetry breaking perturbation such as the Rashba spin-orbit interaction and the magnetic field, 
and are protected completely in the half infinite system.
Finally, existing semiconductor technologies enable junctions between the quantum spin Hall insulator and another materials. 
In addition, the Chern number is tunable by gating the semiconductor. 
This paper provides an alternative definition of the QSHS.

\begin{figure}[htbp]
 \includegraphics[width=85mm]{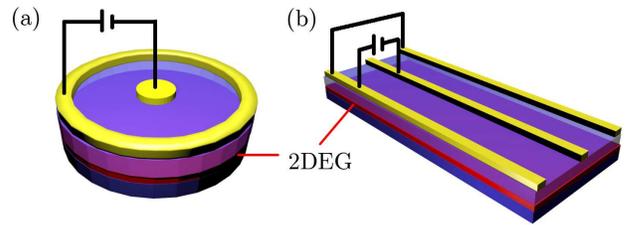}
\caption{The schematic pictures of the system under consideration.
The two-dimensional electron gas (2DEG) is arranged in the quantum well and subjected to the electrostatic potential produced by the gate on the quantum well.
In (a), the electrostatic potential is isotropic and plays a role of the symmetric gauge.
The potential in (b) is symmetric about the center line of the bar geometry and produces the effective Landau gauge in 2DEG.
 }\label{fig1}
\end{figure}
We consider the two-dimensional electric gas (2DEG) described by the single band model 
 in the presence of time-reversal invariant spin-orbit interaction.
The Hamiltonian reads
\begin{align}
\mathcal{H}=\left(\frac{\boldsymbol{p}^2}{2m}+U_0(\boldsymbol{r})\right)\sigma^0
+b\boldsymbol{\sigma}\cdot(\boldsymbol{p}\times\nabla U_0(\boldsymbol{r}))
\end{align}
where $\boldsymbol{p}=(p_x,p_y)$ is two-dimensional momentum and $U_0(\boldsymbol{r})$ is the external electric potential 
applied onto 2DEG in Fig. \ref{fig1}. 
The $2\times2$ matrix $\sigma^\mu$ represents the identity matrix for $\mu=0$ and the Pauli matrices for $\mu=x,y,z$ in spin space. 
The third term is the spin-orbit interaction with the coupling constant $b\hbar$. 
We consider two types of the external static potential $U_0(\boldsymbol{r})$.
The first one $U_0^{\mathrm{S}}(\boldsymbol{r})$ has the potential minimum at a single point in two-dimensional space $(x,y)$ 
and increases proportionally to the square of the distance from the minimum point.
The second one $U_0^{\mathrm{L}}(\boldsymbol{r})$ has the potential minimum along a line in two-dimensional space $(x,y)$ 
and increases proportionally to the square of the distance from the minimum line.
The potentials can be described by
\begin{align}
U_0^{\mathrm{S}}(\boldsymbol{r})=\frac{u_0}{2}(x^2+y^2),
\end{align}
and 
\begin{align}
U_0^{\mathrm{L}}(\boldsymbol{r})=u_0y^2.\label{landau}
\end{align}
We assume that the 2DEG staying at $z=0$ is thin enough in the $z$ direction so that the mirror symmetry $z\rightarrow-z$ 
holds locally at $z=0$. 
In such case, the expectation values of the momentum and the gradient of the electrostatic potential in the $z$ direction 
vanish as $\langle p_z\rangle=0$ and $\langle \partial_z U\rangle=0$. 
Thus the spin-orbit interaction has only terms proportional to $\sigma^z$ as represented by
\begin{align*}
\frac{e}{m}\boldsymbol{A}^\alpha(\boldsymbol{r})\cdot\boldsymbol{p}\sigma^z,
\end{align*}
with
\begin{align}
\boldsymbol{A}^\alpha(\boldsymbol{r})=\frac{mb}{e}\boldsymbol{e}_z\times\nabla U_0^\alpha (\boldsymbol{r}), \label{adef}
\end{align}
for $\alpha=$S and L, where $\boldsymbol{e}_z$ is the unit vector along $z$ axis.
With the symmetric electric potential, the Hamiltonian of the electric state is
\begin{align}
\mathcal{H}=&\frac{\boldsymbol{\Pi}^2}{2m}+U,\label{Hamiltonian}
\end{align}
where $U=u\sigma^0(x^2+y^2)$ is the scaler quadratic potential with $u=u_0-mb^2{u_0}^2/2$
and $\boldsymbol{\Pi}$ is the matrix including the momentum as
\begin{align}
\boldsymbol{\Pi}=&\boldsymbol{p}\sigma^0+e\boldsymbol{A}^\alpha(\boldsymbol{r})\sigma^z. \label{pidef}
\end{align}
The matrix $e\boldsymbol{A}^\alpha(\boldsymbol{r})$ plays the similar role of the gauge vector in the usual quantum Hall states. 
We can also introduce the "magnetic field matrix" $\tilde{\boldsymbol{B}}$ by
\begin{align}
\tilde{\boldsymbol{B}}=\nabla\times\boldsymbol{A}^\alpha(\boldsymbol{r})\sigma^z.
\end{align}
In both cases $\alpha=\mathrm{S}$ and $\mathrm{L}$, the effective magnetic field matrix has same form of
\begin{align}
\tilde{\boldsymbol{B}}=(0,0,B\sigma^z)\label{effectiveB},
\end{align}
with $B=2mub/e$. In what follows, 
we call $\boldsymbol{A}_{\mathrm{S}}\sigma^z$ and $\boldsymbol{A}_{\mathrm{L}}\sigma^z$ "symmetric" and "Landau" gauge matrix respectively, 
because of their forms $\boldsymbol{A}_{\mathrm{S}}=(B/2)(y,-x)$ and $\boldsymbol{A}_{\mathrm{L}}=B(y,0)$.
In fact, the uniform magnetic field along $z$ axis is typically represented by two gauge vectors: 
the symmetric gauge $(B/2)(y,-x)$ and the Landau gauge $B(y,0)$.
However, the gauge matrix in this paper and the usual gauge vector have the different property under the 
 time-reversal operation $\mathcal{T}=i\sigma^y\mathcal{K}$. 
The gauge matrix $\boldsymbol{A}^\alpha\sigma^z$ is antisymmetric under the time-reversal operation,
\begin{align}
\mathcal{T}\boldsymbol{A}^\alpha\sigma^z\mathcal{T}^\dagger=-\boldsymbol{A}^\alpha\sigma^z.
\end{align}
The Hamiltonian of the electric system is time-reversal invariant
\begin{align}
\mathcal{T}\mathcal{H}(-\boldsymbol{p})\mathcal{T}^\dagger=\mathcal{H}(\boldsymbol{p}),
\end{align} 
because the gauge matrix and the momentum are transformed in the same way under the time-reversal operation. 
The ground state of the Hamiltonian in Eq. (\ref{Hamiltonian}) is the quantum spin Hall state
because the diagonal parts of the Hamiltonian $\mathcal{H}=\mathrm{diag}[h_+,h_-]$ are 
nothing other than the quantum Hall Hamiltonian with opposite magnetic field,
\begin{align}
h_\pm=\frac{\left(\boldsymbol{p}\pm e\boldsymbol{A}^\alpha(\boldsymbol{r})\right)^2}{2m}+U.\label{QH}
\end{align}
where the spin of the two electric states is perfectly polarized inversely.
The two quantum Hall blocked sectors host two gapless edge states with opposite spin.
The scaler quadratic potential $U$ in Eq.~(\ref{Hamiltonian}) broadens the density of states arranged from perfectly degenerated Landau levels~\cite{Joynt1984}.
However, it is also possible to reduce $u$ by tuning $u_0$ to $2/mb$, which realizes the well degenerated Landau levels.
When the chemical potential lies between the Landau levels,
2DEG becomes the quantum spin Hall insulator. 
Our scenario for realizing QSHS does not need the inverted band structure.

To topologically classify the quantum Landau spin Hall states, we introduce the spin Chern number\cite{Sheng2006} defined in the similar 
way in the Quantum Hall state\cite{Thouless1982,Qi2008}.
The spin Chern number is calculated as 
\begin{align}
C_s=&\int d\boldsymbol{p} (\nabla_{\boldsymbol{p}}\times \boldsymbol{a}(\boldsymbol{p}))_z\\
\boldsymbol{a}(\boldsymbol{p})=&-\frac{i}{2}{\sum_{n,s}}'\langle n\;s\boldsymbol{p}|\sigma^z\nabla_{\boldsymbol{p}}|n\;s\boldsymbol{p}\rangle,
\end{align}
where ${\sum}'$ means the summation of the electric states in the occupied bands $n$ and with spin $s$. 
The Berry's curvature $\boldsymbol{a}(\boldsymbol{p})$ is represented by 
\begin{align}
\boldsymbol{a}(\boldsymbol{p})=\frac{1}{2}\left(\boldsymbol{a}_\uparrow(\boldsymbol{p})-\boldsymbol{a}_\downarrow(\boldsymbol{p})\right),
\end{align}
with
\begin{align}
\boldsymbol{a}_s(\boldsymbol{p})=&-i{\sum_{n}}'\langle n\;s\boldsymbol{p}|\sigma^z\nabla_{\boldsymbol{p}}|n\;s\boldsymbol{p}\rangle.\nonumber
\end{align}
Because of the time-reversal symmetry, the spin Chern number for one spin direction must be equal to that in the other spin direction.
As a result, the spin Chern number is defined by the Chern number of the up spin electric states,
\begin{align}
C_s=&\int d\boldsymbol{p} (\nabla_{\boldsymbol{p}}\times \boldsymbol{a}_\uparrow(\boldsymbol{p}))_z,
\end{align}
which is an integer number $\mathbb{Z}$.
Therefore the quantum Hall insulator has the 2$\mathbb{Z}$ conducting channels at its edge where 
$\mathbb{Z}$ channels for the spin up states and $\mathbb{Z}$ channels for the spin down one carry the electric current 
opposite direction to each other. As a consequence, they carry the spin current. 
The ordinary $\mathbb{Z}_2$ quantum spin Hall insulators, on the other hand, host only one edge channel for each spin direction.
We also conclude that the spin Hall conductivity of the quantum Landau spin Hall insulator is quantized by the spin Chern number in the 
absence of spin mixing potentials
\begin{align}
G^s_{xy}=s\frac{e}{h}C_s,
\end{align}
where we take the charge for spin by $s$.

\begin{figure}[htbp]
 \includegraphics[width=85mm]{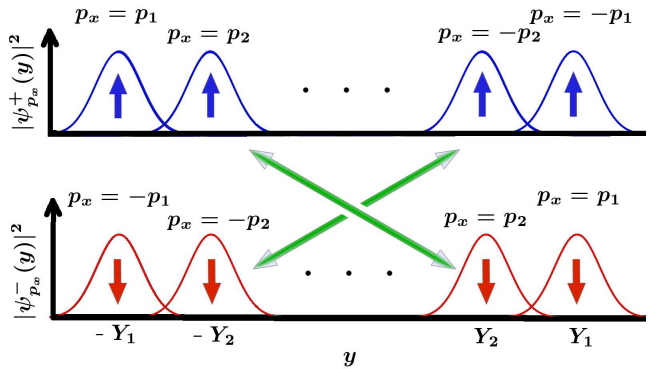}
\caption{The schematic picture of the wave functions localizing at $Y_i=\pm p_i\ell^2/\hbar$ with the eigen momentum $p_x=p_i$ under the Landau gauge.
Here $\psi^+$ and $\psi^-$ are the wave function for up-spin and down-spin states respectively.
The upper and lower series of the eigenstates have up and down spin respectively. The crossing arrows are 
the non-local interactions mixing the states with opposite spin, 
e.g. the Rashba spin-orbit interaction and the uniform Zeeman field.
 }\label{fig2}
\end{figure}

In the following, we discuss the stability of the QLSHS under various perturbations by using the Landau gauge
in Eq.~(\ref{landau}) and 2DEG as shown in Fig.~1(b).
The wave function of eigenstates with the momentum $p_x$ are represented by 
\begin{align}
\Psi_{n,p_x}(x,y)=\frac{e^{ip_xx}}{\sqrt{L}}\left\{c_+\begin{pmatrix}
\psi_{n,p_x}^+(y)\\0
\end{pmatrix}+
c_-\begin{pmatrix}
0\\
\psi_{n,p_x}^-(y)
\end{pmatrix}\right\},
\end{align}
with
\begin{align}
\psi_{n,p_x}^\pm(y)=\left(\frac{e^{-(y\pm Y)^2/\ell^2}}{\sqrt{\pi}2^nn!\ell}\right)^{1/2}H_n\left(\frac{y\pm Y}{\ell}\right)\label{wavefunction},
\end{align}
where $\psi_{n,p_x}^\pm$ is localized at $y=\pm Y$ with $Y=-p_x\ell^2/\hbar$ and $\ell=\sqrt{\hbar/eB}$ being the spatial width of the edge channel.
Since the energy of QHS in the each blocked spin sector is characterized by Landau levels, the edge states 
play a dominant role in the quantization of the spin Hall and electric conductivities~\cite{MacDonald1984,Kane2005b}.
Therefore we consider effects of various perturbations on the electric states near the edges under the condition of $\ell\ll Y$.
We consider the perturbed Hamiltonian represented by
\begin{align}
\mathcal{H}'=&d_0\sigma^0+d_\mu\sigma^\mu,
\end{align}
where the index which appears twice in a single term means the summation for $\mu=x,y,z$.
In the presence of the time-reversal invariance, $d_0$ and $d_\mu$ are symmetric and antisymmetric under $\boldsymbol{p}\rightarrow-\boldsymbol{p}$, respectively.
The spin independent disorder $d_0\sigma^0$ 
does not affect the QLSHSs because the quantum Hall state in each blocked spin sector in Eq. (\ref{Hamiltonian}) is robust against the potential disorder.
The Rashba spin-orbit interaction due to the potential gradient along $z$ axis
breaks the mirror symmetry and is represented by 
\begin{align}
\mathcal{H}_{\mathrm{R}}=&\lambda(p_x\sigma^y-p_y\sigma^x).\label{Rashba}
\end{align}
Such perturbation mixes the two spin sectors. 
The interaction $p_x\sigma^y$ mixes the opposite spin state with the same momenta $p_x$ and the same Landau index. However such edge channels 
are localized opposite edges to each other and are spatially separated by $2Y$
as represented by $\psi_{n,p_x}^+$ and $\psi_{n,p_x}^-$ in  
Eq. (\ref{wavefunction}). 
In Fig. \ref{fig2}, we schematically illustrate these wave functions,
where $\psi^+$ in upper figure and $\psi^-$ in the lower one represent the wave functions for spin-up state and these for spin-down one, respectively.
A spin-up state with $p_x>0$ and a spin-down one with $p_x<0$ are localized at the same edge.
On the other hand, a spin-up state and a spin-down one having the same $p_x$ are localized at the opposite edge to each other.
As a consequence, the transition probability becomes small value proportional to $\exp[-(Y/\ell)^2]$.
Thus the edge states are protected perfectly in the half-infinite system. 
This is a key feature which protects QLSHS under the various perturbation as discussed below.
The interaction $p_y\sigma^x$ conserves $p_x$ but mixes the eigenstates of different Landau level because of
\begin{align}
p_y\psi^\pm_{n,p_x}=\frac{y-Y}{2\ell^2}\psi^\pm_{n,p_x}-\frac{\sqrt{2(n+1)}}{\ell}\psi^\pm_{n+1,p_x}.
\end{align}
Scattering due to the first and second terms are negligible by the same reason as discussed in the $p_x\sigma^y$ term.
When the Rashba spin-orbit interaction is weak enough to satisfy $\hbar\lambda/B\ell$,
the electric states remain insulating in the bulk and helical at the edges.

Even in the absence of the time-reversal symmetry, the edge states of the QLSHS are stable under $\ell\ll Y$ in some cases.
Here we discuss the uniform Zeeman field described by $\boldsymbol{B}_{\mathrm{ext}}\cdot\boldsymbol{\sigma}$.
The uniform Zeeman field preserves the momentum $p_x$ and the Landau level $n$.
The Zeeman field in the $z$ direction is spin diagonal and only shifts the energies of the two quantum Hall Hamiltonians of Eq.~(\ref{QH}).
Effects of the Zeeman field in the $xy$ plane are also negligible for $\ell\ll Y$ by the same reason as discussed in the $p_x\sigma^y$ term.
Next we consider uniform magnetic field applied onto 2DEG represented by $\boldsymbol{B}_{\mathrm{ext}} = \nabla \times \boldsymbol{A}_{\mathrm{ext}}$ 
which shift the momentum as $\boldsymbol{p}\rightarrow \boldsymbol{p}-e\boldsymbol{A}_{\mathrm{ext}}$.
As a result, the gauge matrix in $\boldsymbol{\Pi}$ at Eq.~(\ref{pidef}) changes to $\boldsymbol{A}_{\mathrm{total}}= \boldsymbol{A}^L\sigma^z +
\boldsymbol{A}_{\mathrm{ext}}\sigma^0$.
The real magnetic field only modifies the amplitude of the gauge matrix depending on spin direction.
Therefore, the edge states of the QLSHS is stable even in the presence of uniform magnetic fiels with $B_{\mathrm{ext}} < B$.
However, magnetic impurities break the quantization of the spin Hall and electric conductivities in the QLSHS because they mixes 
the opposite spin edge channels localizing the same edge.

Finally, we estimate the strength of the applied electrostatic potential $u_0$ to realize the QLSHS.
The energy gap of the Landau levels $\Delta E_c$ can be represented by
\begin{align}
\Delta E_c=\frac{\hbar e B}{m}=2\hbar bu_0.
\end{align}
For instance, in the quantum well of In$_x$Ga$_{1-x}$As sandwiched by InP\cite{Engels1997}, the Rashba spin-orbit coupling 
constant $\alpha=b\hbar\langle E_z\rangle$ is estimated to be $10^{-11}[\mathrm{eVm}]$, where $\langle E_z\rangle$ is the electric field perpendicular to the quantum well.
In experiments, the electric field is typically estimated to be $10^{6}[\mathrm{V/m}]$. 
The coupling constant of spin-orbit interaction results in $b\hbar\simeq10^{-17}[\mathrm{m}^{2}]$.
To realize the QLSHS with $\Delta E_c=10[\mathrm{meV}]$ in a quantum well with the sample width of $0.1[\mu\mathrm{m}]$, the voltage between 
the edge and the point of the potential minimum is required to be $5[\mathrm{V}]$.

In summary, we have studied the two-dimensional electric states on the single-band semiconductor 
under the quadratic electrostatic potential. 
The strong enough spin-orbit interaction playing a role of the effective gauge vector causes 
the Landau level quantization of the two-dimensional electronic states. As a consequence, 
the electronic states become the two-dimensional quantum spin Hall states showing the property of 
the quantized spin Hall conductivity. 
Under the effective Landau gauge, we have shown that 
the quantum Landau spin Hall states are robust under various perturbations such as the Rashba spin-orbit 
interaction and magnetic fields.

This work was supported by 
the "Topological Quantum Phenomena" (No. 22103002) Grant-in Aid for 
Scientific Research on Innovative Areas from the Ministry of Education, 
Culture, Sports, Science and Technology (MEXT) of Japan.
\bibliography{TI}

\end{document}